\documentclass[twocolumn,superscriptaddress,amsmath, amssymb, amsfonts,preprintnumbers,aps,prd,longbibliography,nofootinbib]{revtex4-1}

\usepackage{scalerel}
\usepackage{xcolor}
\usepackage{amsmath,amsfonts,amssymb}
\usepackage[small,bf,hang]{caption}
\usepackage{slashed}
\usepackage{latexsym,epsfig}
\usepackage{dsfont}
\usepackage{arydshln}
\usepackage{extarrows}
\usepackage{hyperref}
\usepackage{array}

\def\moth{\mathsurround=0pt}
\newdimen\zo \zo=0pt

\def\tick{\leaders\hrule height 0.5ex depth 0pt \hskip 0.5pt}
\def\upboxfill{$\moth \setbox\zo\hbox{\tick}%
  \hskip 3pt\hbox to 0pt{$\tick$\hss}\hrulefill \hbox to 7.5pt{$\tick$\hss}$}

\def\dtick{\leaders\hrule height .34pt depth 0.5ex \hskip 0.5pt}
\def\downboxfill{$\moth \setbox\zo\hbox{\dtick}%
  \hskip 2pt\hbox to 0pt{$\dtick$\hss}\hrulefill \hbox to 2pt{$\dtick$\hss}$}

\def\bec{\begin{center}}
\def\ec{\end{center}}

\def\nn{\nonumber}

\def\be{\begin{equation}}
\def\ee{\end{equation}}
\def\bea{\begin{eqnarray}}
\def\eea{\end{eqnarray}}
\def\ba{\begin{array}}
\def\ea{\end{array}}

\begin{document}

\title{Curvatures and Non-metricities in the Non-Relativistic Limit of Bosonic Supergravity}

\author{Eric Lescano} 
\email{eric.lescano@uwr.edu.pl}
\affiliation{Institute for Theoretical Physics (IFT), University of Wroclaw, \\
pl. Maxa Borna 9, 50-204 Wroclaw,
Poland}

\begin{abstract}
We construct a metric-like formulation of the non-relativistic (NR) limit of bosonic supergravity at the Lagrangian level. This formulation is particularly useful for decomposing relativistic tensors, such as powers of the Riemann tensor, in a manifest covariant form with respect to infinitesimal diffeomorphisms. The construction is purely geometrical and is based on a torsionless connection, mimicking the construction of the relativistic theory. The formulation contains non-vanishing non-metricities, which are associated with the gravitational fields of the theory ($\tau_{\mu\nu}$, $h_{\mu\nu}$, $\tau^{\mu\nu}$, $h^{\mu\nu}$). The non-metricities are fixed by requiring compatibility with the relativistic metric, before taking the NR expansion. In this sense, they can be interpreted as intrinsic non-metricities. We provide a fully covariant decomposition of the relativistic Riemann tensor, Ricci tensor, and scalar curvature. Our results establish an equivalence between the vielbein approach  of string Newton--Cartan geometry at the level of the Lagrangian and the proposed construction. We also discuss potential applications, including a pure metric rewriting of the two-derivative finite bosonic supergravity Lagrangian under the NR limit, a powerful simplification in deriving NR bosonic $\alpha'$-corrections and extensions to more general $f(R,Q)$ Newton--Cartan geometries.
\end{abstract}

\maketitle

\section{Introduction}

In recent years, non-relativistic (NR) limits of string theory \cite{NR1}-\cite{NR3} and supergravity \cite{NRST1}-\cite{NRST13} have attracted considerable attention, motivated by both conceptual developments and potential phenomenological applications. In particular, the systematic study of the supergravity limit of NR string theory has led to the emergence of Newton--Cartan and string Newton--Cartan geometries as the natural geometric frameworks underlying these limits (for reviews and complementary introductions, see \cite{Review}-\cite{Review3} and references therein).

While the NR limit of NS-NS gravity has been deeply studied in \cite{NSNS}, where the authors have addressed the construction in the vielbein formalism of NR theory. In this work we will reformulate this same setup, but using a purely metric approach, where the full Lorentz symmetry is manifest from the starting point. In other words, we will address NSNS-gravity in its metric formulation directly from the relativistic theory, and then we will study extensions which keep this structure. This way of avoiding the vielbein formulation will be extremely useful to address higher-derivative contributions in its NR limit, which can be written entirely with curvatures, like powers of the Riemann tensor. In this sense, one can avoid the use of the spin connection, and work solely with an affine connection. Our canonical example here is the Metsaev and Tseytlin formulation of the four-derivative corrections of bosonic string theory \cite{MetsaevTseytlin}. 

Let us begin by considering the universal bosonic NS--NS supergravity, whose relativistic field content consists of a spacetime metric $\hat g_{\mu\nu}$, a Kalb--Ramond two-form $\hat B_{\mu\nu}$ and a dilaton $\hat\phi$. A controlled NR limit can be obtained \cite{NSNS} by introducing an explicit expansion in powers of $c$,
\bea
    \hat{g}_{\mu \nu} & = & c^2 \tau_{\mu \nu} + h_{\mu \nu} \, , \label{metricexpansionintro} \\
    \hat{g}^{\mu \nu} & = & \frac{1}{c^2} \tau^{\mu \nu} + h^{\mu \nu} \, , \\
    \hat{B}_{\mu \nu} & = & - c^2 c_{\mu \nu} + b_{\mu \nu} \, , \\
    \hat{\phi} & = & \ln(c) + \varphi \, ,
\eea
where $\mu,\nu=0,\dots,25$. The longitudinal part can be conveniently parametrized in terms of the vector $\tau^{\mu}$ and the 1-form $\tau_{\mu}$ \footnote{Some of our results will be presented in terms of these objects, since they help to simplify long expressions.}, 
\bea
\tau_{\mu \nu} = \tau_{\mu} \tau_{\nu} \, , \\
\tau^{\mu \nu} = \tau^{\mu} \tau^{\nu} \, .
\eea
This vectorial decomposition is introduced as a convenient ansatz that allows the divergent contributions of the Ricci scalar and the Kalb--Ramond sector to be identified explicitly and written in a compact form. It should not be interpreted as the longitudinal vielbein decomposition of the standard string Newton--Cartan formulation. Throughout this work we employ a purely metric formulation, in which the fundamental variables are the longitudinal and transverse metric components, rather than the vielbein \footnote{The $SO(1,1)$ indices associated with the latter appear only when discussing the constraints that the boost transformations impose over the non-metricities.}.

The Newton-Cartan variables satisfy the defining relations
\bea
h^{\mu \nu} \tau_{\nu \rho} & = & h^{\mu \nu} c_{\nu \rho} = 0 \, , \\
h_{\mu \nu} \tau^{\nu \rho} & = & 0 \, , \\ 
h_{\mu \nu} h^{\nu \rho} & = & \delta_{\mu}^{\rho} - \tau_{\mu \nu} \tau^{\nu \rho} \, , \label{delta}
\eea

Although individual geometric quantities constructed from the relativistic Levi--Civita connection diverge in the limit $c\to\infty$, the bosonic NS--NS action
\bea
S = \int d^{26}x \sqrt{-\hat g} e^{-2 \hat \phi} 
\Big( \hat R + 4 \partial_{\mu} \hat \phi \partial^{\mu} \hat \phi 
- \frac{1}{12} \hat H_{\mu \nu \rho} \hat H^{\mu \nu \rho}\Big) \, ,
\label{Sbosintro}
\eea
with $\hat H_{\mu\nu\rho}=3\partial_{[\mu}\hat B_{\nu\rho]}$, remains finite due to a non-trivial cancellation between divergent contributions arising from the Ricci scalar and the Kalb--Ramond sector (in vielbein formalims $c_{\mu \nu}=\tau_{\mu}{}^{a} \tau_{\nu}{}^{b} \epsilon_{a b}$ ensures the cancellation, with $a,b=0,1$ the transversal flat directions). This mechanism underlies the construction of various non-relativistic string supergravity theories, including heterotic formulations \cite{BandR}-\cite{Sigmaheterotic}, supersymmetric extensions \cite{NRST11}-\cite{NRST12} and Double Field Theory constructions \cite{DFT1}-\cite{DFT4}, as well as their formulation within non-Riemannian double geometries \cite{NRDFT1}-\cite{NRDFT9} and their NR limits \cite{EandD}.

Motivated by these developments, increasing attention has recently been devoted to understanding higher-derivative corrections in the NR limit \cite{Higher-Derivative1}-\cite{Higher-Derivative4}. At the relativistic level, many of these corrections are naturally written in terms of covariant tensors constructed from the Levi--Civita connection. In order to study their NR limit while preserving covariance under spacetime diffeomorphisms, it is useful to express these objects in terms of curvatures and covariant derivatives adapted to the NR geometric variables appearing in the expansion \eqref{metricexpansionintro}. 

Previous formulations of the NR limit of NS--NS gravity provide a consistent description of the dynamics in terms of Newton--Cartan geometry and connections that naturally accommodate the local boost, $SO(1,1)$ and transverse rotation symmetries of the theory \cite{NSNS}. While this framework is well suited for describing the two-derivative dynamics of NR supergravity certain applications, in particular the systematic decomposition of relativistic curvature tensors appearing in higher-derivative corrections, benefit from a pure metric formulation. In such a formulation, the relativistic curvature invariants could be reorganized in a covariant manner in terms of their different contributions in powers of $c$ without need of using the vielbein decomposition or spin connections (see also \cite{new} for a metric formalism in p-brane Galilean geometries).

The main goal of this work is to provide such a formulation. We construct a torsionless affine connection adapted to the NR metric variables $\tau_{\mu\nu}$ and $h_{\mu\nu}$ and use it to rewrite relativistic geometric quantities in a manifestly covariant form under spacetime diffeomorphisms. Importantly, the resulting connection is not metric compatible with respect to the fundamental NR fields. Instead, the covariant derivatives of the fundamental fields are controlled by a set of non-metricity tensors that arise naturally from expanding the relativistic metric compatibility condition. These non-metricities are uniquely determined by requiring consistency with the relativistic Levi--Civita structure (compatibility with the relativistic metric).

This construction allows relativistic curvature tensors to be systematically reorganized into NR geometric invariants. In particular, we show that the full bosonic supergravity action can be rewritten in terms of covariant NR quantities, with all contributions remaining finite and well defined in the limit $c\rightarrow\infty$.

The purpose of this paper is threefold. First, we present an explicit construction of a torsionless affine connection and the associated non-metricity tensors within the Newton--Cartan framework. Second, we compute the corresponding curvature tensors and demonstrate how relativistic curvature invariants decompose into NR geometric objects in a covariant manner. Third, we illustrate several applications of the formalism. These include the derivation of the complete NR two-derivative bosonic supergravity Lagrangian in a manifestly covariant form, the analysis of the finite form of the Metsaev-Tseytlin alpha'-contributions, and the construction of more general Newton--Cartan-inspired theories based on alternative choices of non-metricity that extend beyond the specific structure inherited from bosonic supergravity.

Overall, our results provide a geometric framework for the NR limit of bosonic supergravity that parallels the role played by the Levi--Civita connection in relativistic theories, while naturally incorporating non-metricity effects. We expect this formalism to provide a useful starting point for the systematic study of higher-derivative corrections and duality-covariant formulations in the non-relativistic limit of string theory.

\section{Pure metric formalism and the affine connection}
We start by recalling the transformation rule of the fundamental fields with respect to infinitesimal diffeomorphisms,
\bea
\delta_{\xi} \tau_{\mu \nu} & = &  \xi^{\rho} \partial_{\rho}{\tau_{\mu \nu}} + 2 \partial_{(\mu|}{\xi^{\rho}} \tau_{\rho |\nu)} \, , \nn \\
\delta_{\xi} \tau^{\mu \nu} & = & \xi^{\rho} \partial_{\rho} \tau^{\mu \nu} - 2 \partial_{\rho}{\xi^{(\mu}} \tau^{\rho \nu)} \, , 
\nn \\
\delta_{\xi} h_{\mu \nu} & = & \xi^{\rho} \partial_{\rho} h_{\mu \nu} + 2 \partial_{(\mu|} \xi^{\rho} h_{|\nu) \rho} \, , \nn \\
\delta_{\xi} h^{\mu \nu} & = & \xi^{\rho} \partial_{\rho}{h^{\mu \nu}} – 2 \partial_{\rho}{\xi^{(\mu}} h^{ \nu) \rho}  \, .
\eea

The torsionless connection is given by
\bea
&& \Gamma_{\mu \nu}^{\rho}(\tau,h) = \frac12 h^{\rho \sigma} (2 \partial_{(\mu} h_{\nu) \sigma} - \partial_{\sigma} h_{\mu \nu}) \nn \\ && + \frac12 \tau^{\rho \sigma} (2 \partial_{(\mu} \tau_{\nu) \sigma} - \partial_{\sigma}\tau_{\mu \nu}) \, ,
\eea
which is the $c^0$-contribution coming from the relativistic connection, 
\bea
\hat \Gamma_{\mu \nu}^{\rho}(\hat g) = c^2 \hat \Gamma_{\mu \nu}^{(2)\rho}(\tau,h) + \Gamma_{\mu \nu}^{\rho}(\tau,h) + \frac{1}{c^2} \hat \Gamma_{\mu \nu}^{(-2)\rho}(\tau,h) \, ,
\eea
preserving the non-tensorial transformation rule. In other words, $\hat \Gamma_{\mu \nu}^{(2)\rho}(\tau,h)$ and $\hat \Gamma_{\mu \nu}^{(-2)\rho}(\tau,h)$ transform as tensors, so we can shift away their contributions and consider only $\Gamma_{\mu \nu}^{\rho}(\tau.h)$ as the non-relativistic connection. This result only requires the notion of the longitudinal part of the metric, $\tau_{\mu \nu}$, and the transversal one, $h_{\mu \nu}$ (and similarly for the inverse metric). The choice of the connection is not unique, and it can be (re-)shifted by covariant contributions.

The covariant derivatives are therefore defined as
\bea
\nabla_{\mu} \tau_{\nu \sigma} & = & \partial_{\mu} \tau_{\nu \sigma} - 2 \Gamma_{\mu (\nu|}^{\rho} \tau_{\rho |\sigma)} \, , \\
\nabla_{\mu} \tau^{\nu \rho}& = & \partial_{\mu} \tau^{\nu \rho} + 2 \Gamma_{\mu \sigma}^{(\nu|} \tau^{\sigma |\rho)} \, , \\
\nabla_{\mu} h_{\nu \rho} & = & \partial_{\mu} h_{\nu \rho} - 2 \Gamma_{\mu (\nu}^{\sigma} h_{\rho) \sigma} \, , \\
\nabla_{\mu} h^{\nu \rho} & = & \partial_{\mu} h^{\nu \rho} + 2 \Gamma_{\mu \sigma}^{(\nu} h^{\rho) \sigma} \, . 
\eea
The Riemann tensor can be easily computed from the commutator of the covariant derivatives acting on an arbitrary vector $v^{\mu}$,
\bea
[\nabla_{\mu},\nabla_{\nu}] v^{\rho} = R^{\rho}{}_{\epsilon \mu \nu} v^{\epsilon} \, , 
\eea
giving the usual expression
\bea
R^{\rho}{}_{\epsilon \mu \nu}(\tau,h) = 2 \partial_{[\mu} \Gamma_{\nu] \epsilon}^{\rho} + 2 \Gamma_{[\mu| \alpha}^{\rho} \Gamma_{|\nu] \epsilon}^{\alpha} \, .
\eea

\section{Intrinsic non-metricities}
The non-metricities play a fundamental role in the theory, since they guarantee that the NR limit can be taken from the relativistic theory,
\bea
\nabla_{\mu} \tau_{\nu \rho} & = & Q^{(\tau)}_{\mu \nu \rho} \, , \\
\nabla_{\mu} \tau^{\nu \rho} & = & Q^{(\tau^{-1})}_{\mu}{}^{\nu \rho} \, ,  \\
\nabla_{\mu} h_{\nu \rho} & = & Q^{(h)}_{\mu \nu \rho} \, , \\
\nabla_{\mu} h^{\nu \rho} & = & Q^{(h^{-1})}_{\mu}{}^{\nu \rho} \, . 
\eea
The fixed/intrinsic values of the non-metricities come from the metric conditions,
\bea
\label{metriccond1}
\hat \nabla_{\mu} \hat g_{\nu \rho} & = & 0 \, , \\
\hat \nabla_{\mu} \hat g^{\nu \rho} & = & 0 \, ,
\label{metriccond2}
\eea
where $\hat \nabla$ is the usual covariant derivative with respect to the relativistic Levi--Civita connection. Therefore, after performing the NR expansion, one needs the following non-metricity combinations,
\bea
\label{Q1}
Q^{(\tau)}_{\mu \nu \rho} & = & 2 \Gamma^{(2)\sigma}{}_{\mu (\nu|} h_{\sigma|\rho)}  \\
Q^{(\tau^{-1})}_{\rho}{}^{\mu \nu} & = & - \tau^{(\mu \alpha} (2 \partial_{(\rho} h_{\sigma) \alpha} - \partial_{\alpha} h_{\rho \sigma}) h^{\sigma \nu)}  \\
Q^{(h)}_{\mu \nu \rho} & = & 2 \Gamma^{(-2)\sigma}{}_{\mu (\nu|} \tau_{\sigma|\rho)}  \\
Q^{(h^{-1})}_{\rho}{}^{\mu \nu} & = & - h^{(\mu \alpha} (2 \partial_{(\rho} \tau_{\sigma) \alpha} - \partial_{\alpha} \tau_{\rho \sigma}) \tau^{\sigma \nu)} \, , 
\label{Q4}
\eea
where
\bea
\Gamma^{(2)\sigma}_{\mu \nu} = \frac12 h^{\alpha \sigma} (\nabla_{\mu} \tau_{\nu \alpha} + \nabla_{\nu} \tau_{\mu \alpha} - \nabla_{\alpha} \tau_{\mu \nu}), \\
\Gamma^{(-2)\sigma}_{\mu \nu} = \frac12 \tau^{\alpha \sigma} (\nabla_{\mu} h_{\nu \alpha} + \nabla_{\nu} h_{\mu \alpha} - \nabla_{\alpha} h_{\mu \nu}), \, .
\eea

The NR non-metricities are not boost invariants. We recall that the boost transformations are given by \cite{NSNS},
\bea
\label{transfzeroa}
\delta_{\lambda} h^{\mu \nu} & = & 0 \, , \\
\delta_{\lambda} h_{\mu \nu} & = & - 2 \lambda_{a a'} e_{(\mu}{}^{a'} \tau_{\nu)}{}^{a}  \, , \\
\delta_{\lambda} \tau^{\mu \nu} & = & 2 \lambda_{a}{}^{a'} e^{(\mu}{}_{a'} \tau^{\nu) a}  \, ,  \\
\delta_{\lambda} \tau_{\mu \nu} & = & 0 \, ,  \\
\delta_{\lambda} b_{\mu \nu} & = & - 2 \epsilon_{a b} \lambda^{a}{}_{a'} \tau_{[\mu}{}^{b} e_{\nu]}{}^{a'} \, , \\ 
\delta_{\lambda} c_{\mu \nu} & = & 0 \, ,
\label{transfzerob}
\eea
where $\lambda_{a a'}$ is an arbitrary boost parameter. Here we are momentarily including the $SO(1,1)$ index, as discussed in the previous section, to explicitly show that in this geometry, $Q=0$ is not possible for all the metricities. In other words, some of the non-metricities cannot be set to zero without compromising the boost symmetry. We will return to this point in the discussion section, since one might be interested in exploring Newton-Cartan geometries with more arbitrary non-metricities, beyond bosonic supergravity.

\section{Decomposition of the relativistic curvatures}

\subsection{The Riemann tensor}
The relativistic Riemann tensor is given by
\bea
\hat R^{\rho}{}_{\epsilon \mu \nu} & = & c^{4} \hat R^{(4)\rho}{}_{\epsilon \mu \nu} + c^{2} \hat R^{(2)\rho}{}_{\epsilon \mu \nu} + \hat R^{(0)\rho}{}_{\epsilon \mu \nu} \nn \\ && + c^{-2} \hat R^{(-2)\rho}{}_{\epsilon \mu \nu} + c^{-4} \hat R^{(-4)\rho}{}_{\epsilon \mu \nu} \, .
\eea
So far we have just identified part of the covariant contribution inside $\hat R^{(0)\rho}{}_{\epsilon \mu \nu}$, given by $R^{\rho}{}_{\epsilon \mu \nu}$ but other covariant contributions are encoded in $\hat R^{(0)\rho}{}_{\epsilon \mu \nu}$ due to $c^2$- and $c^{-2}$-contributions in the relativistic Levi-Civita connection. All the contributions in the relativistic Riemann tensor can be written in covariant form, and this is what we are going to show now. Let us start with the higher-order contribution,
\bea
\hat R^{(4) \rho}{}_{\epsilon \mu \nu} & = & \frac12 h^{\rho \sigma} h^{\alpha \beta} (\partial_{\alpha} \tau_{[\mu \sigma} - \partial_{\sigma} \tau_{[\mu \alpha}) \nn \\
&& (\partial_{\nu]} \tau_{\epsilon \beta} + \partial_{\epsilon} \tau_{\nu] \beta} - \partial_{\beta} \tau_{\nu] \epsilon}) \, .
\eea
The previous quantity transforms as a tensor, and it can be written in a manifest covariant form using the covariant derivatives,
\bea
\hat R^{(4) \rho}{}_{\epsilon \mu \nu} & = & \frac12 h^{\rho \sigma} h^{\alpha \beta} (\nabla_{\alpha} \tau_{[\mu \sigma} - \nabla_{\sigma} \tau_{[\mu \alpha}) \nn \\
&& (\nabla_{\nu]} \tau_{\epsilon \beta} + \nabla_{\epsilon} \tau_{\nu] \beta} - \nabla_{\beta} \tau_{\nu] \epsilon}) \, .
\eea

The same technique can be implemented for the lowest contribution of the Riemann tensor,
\bea
\hat R^{(-4) \rho}{}_{\epsilon \mu \nu} & = & \frac12 \tau^{\rho \sigma} \tau^{\alpha \beta} (\partial_{\alpha} h_{[\mu \sigma} - \partial_{\sigma} h_{[\mu \alpha}) \nn \\
&& (\partial_{\nu]} h_{\epsilon \beta} + \partial_{\epsilon} h_{\nu] \beta} - \partial_{\beta} h_{\nu] \epsilon})
\eea
which therefore can be written as
\bea
\hat R^{(-4) \rho}{}_{\epsilon \mu \nu} & = & \frac12 \tau^{\rho \sigma} \tau^{\alpha \beta} (\nabla_{\alpha} h_{[\mu \sigma} - \nabla_{\sigma} h_{[\mu \alpha}) \nn \\
&& (\nabla_{\nu]} h_{\epsilon \beta} + \nabla_{\epsilon} h_{\nu] \beta} - \nabla_{\beta} h_{\nu] \epsilon}) \, .
\eea
In these cases we have identified that the structure of $R^{(-4) \rho}{}_{\epsilon \mu \nu}$ is just an interchange of $\tau_{\mu \nu} \leftrightarrow h_{\mu \nu}$ and $h^{\mu \nu} \leftrightarrow \tau^{\mu \nu}$ with respect to the highest order contribution, $R^{(4) \rho}{}_{\epsilon \mu \nu}$.

The $c^2$-contributions of the relativistic Riemann tensor take an important role, since it is part of bosonic supergravity action once one takes the NR limit. These contributions can be written in terms of the covariant derivatives
\bea
\hat R^{(2)\rho}{}_{\epsilon \mu \nu} = 2 \nabla_{[\mu} \Big[ h^{\rho \alpha}( \nabla_{\nu]} \tau_{\epsilon \alpha} + \nabla_{\epsilon} \tau_{\nu] \alpha} - \nabla_{\alpha} \tau_{\nu] \epsilon}) \Big] \, .
\eea
Again, we can interchange $\tau_{\mu \nu} \leftrightarrow h_{\mu \nu}$ and $h^{\mu \nu} \leftrightarrow \tau^{\mu \nu}$ to construct the $c^{(-2)}$-contributions,
\bea
\hat R^{(-2)\rho}{}_{\epsilon \mu \nu} = 2 \nabla_{[\mu} \Big[ \tau^{\rho \alpha}( \nabla_{\nu]} h_{\epsilon \alpha} + \nabla_{\epsilon} h_{\nu] \alpha} - \nabla_{\alpha} h_{\nu] \epsilon}) \Big] \, .
\eea
Finally, we proceed to construct the $c^{0}$-contributions to the relativistic Riemann tensor. This quantity is given by,
\begin{widetext}
\bea
 \hat R^{(0)\rho}{}_{\epsilon \mu \nu} & =  R^{\rho}{}_{\epsilon \mu \nu} + \frac12 h^{\rho \sigma} \tau^{\alpha \beta} (\nabla_{[\mu} \tau_{\alpha \sigma} + \nabla_{\alpha} \tau_{[\mu \sigma} - \nabla_{\sigma} \tau_{[\mu \alpha})  
(\nabla_{\nu]} h_{\epsilon \beta} + \nabla_{\epsilon} h_{\nu] \beta} - \nabla_{\beta} h_{\nu] \epsilon}) 
\nn \\ & \qquad \quad \qquad
+ \frac12 \tau^{\rho \sigma} h^{\alpha \beta} (\nabla_{[\mu} h_{\alpha \sigma} + \nabla_{\alpha} h_{[\mu \sigma} - \nabla_{\sigma} h_{[\mu \alpha})  
(\nabla_{\nu]} \tau_{\epsilon \beta} + \nabla_{\epsilon} \tau_{\nu] \beta} - \nabla_{\beta} \tau_{\nu] \epsilon}) \, .
\eea
In the next section we will study the decomposition of the (relativistic) Ricci tensor in terms of the Newton-Cartan fields, and in covariant form with respect to infinitesimal diffeomorphisms.

\subsection{The Ricci tensor}
The relativistic Ricci tensor is given by
\bea
\hat R_{\epsilon \nu} & = & c^{4} \hat R^{(4)}_{\epsilon \nu} + c^{2} \hat R^{(2)}_{\epsilon \nu} + \hat R^{(0)}_{\epsilon \nu} + c^{-2} \hat R^{(-2)}_{\epsilon \nu} + c^{-4} \hat R^{(-4)}_{\epsilon \nu} \, .
\eea
Using the results of the previous subsection, we can directly write all the contributions in covariant form,
\bea
\hat R^{(4)}_{\epsilon \nu} & = & \frac12 h^{\mu \sigma} h^{\alpha \beta} (\nabla_{\alpha} \tau_{[\mu \sigma} - \nabla_{\sigma} \tau_{[\mu \alpha})
(\nabla_{\nu]} \tau_{\epsilon \beta} + \nabla_{\epsilon} \tau_{\nu] \beta} - \nabla_{\beta} \tau_{\nu] \epsilon}) \, ,
\eea
\bea
\hat R^{(2)}_{\epsilon \nu} = 2 \nabla_{[\mu} \Big[ h^{\mu \alpha}( \nabla_{\nu]} \tau_{\epsilon \alpha} + \nabla_{\epsilon} \tau_{\nu] \alpha} - \nabla_{\alpha} \tau_{\nu] \epsilon}) \Big] \, ,
\eea
\bea
\hat R^{(0)}_{\epsilon \nu} = &  R_{\epsilon \nu}(\tau,h) + \frac12 h^{\mu \sigma} \tau^{\alpha \beta} \nabla_{\alpha} \tau_{[\mu \sigma}
(\nabla_{\nu]} h_{\epsilon \beta} + \nabla_{\epsilon} h_{\nu] \beta} - \nabla_{\beta} h_{\nu] \epsilon}) 
\nn \\ & \qquad \qquad 
+ \frac12 \tau^{\mu \sigma} h^{\alpha \beta} \nabla_{\alpha} h_{[\mu \sigma}  
(\nabla_{\nu]} \tau_{\epsilon \beta} + \nabla_{\epsilon} \tau_{\nu] \beta} - \nabla_{\beta} \tau_{\nu] \epsilon})
\eea
\bea
\hat R^{(-2)}_{\epsilon \nu} = 2 \nabla_{[\mu} \Big[ \tau^{\mu \alpha}( \nabla_{\nu]} h_{\epsilon \alpha} + \nabla_{\epsilon} h_{\nu] \alpha} - \nabla_{\alpha} h_{\nu] \epsilon}) \Big] \, .
\eea
\bea
\hat R^{(-4)}{}_{\epsilon \nu} & = & \frac12 \tau^{\mu \sigma} \tau^{\alpha \beta} (\nabla_{\alpha} h_{[\mu \sigma} - \nabla_{\sigma} h_{[\mu \alpha})(\nabla_{\nu]} h_{\epsilon \beta} + \nabla_{\epsilon} h_{\nu] \beta} - \nabla_{\beta} h_{\nu] \epsilon}) \, .
\eea
\subsection{The Ricci scalar}
The relativistic Ricci scalar is decomposed as
\bea
\hat R & = & c^{2} \hat R^{(2)}  + \hat R^{(0)} + c^{-2} \hat R^{(-2)} + c^{-4} \hat R^{(-4)} \, ,
\eea
where the covariant form of each contribution is given by
\bea
\hat R^{(2)} & = & \hat R^{(2)}_{\mu \nu} h^{\mu \nu} + \hat R^{(4)}_{\mu \nu} \tau^{\mu \nu} \, ,   \\    
\hat R^{(0)} & = & \hat R^{(0)}_{\mu \nu} h^{\mu \nu} + \hat R^{(2)}_{\mu \nu} \tau^{\mu \nu} \, , \\ 
\hat R^{(-2)} & = & \hat R^{(-2)}_{\mu \nu} h^{\mu \nu} + \hat R^{(0)}_{\mu \nu} \tau^{\mu \nu} \, , \\
\hat R^{(-4)} & = & \hat R^{(-4)}_{\mu \nu} h^{\mu \nu} \, .
\eea
Using the results of the previous subsection we can easily see that the right hand side of the previous contributions can be written in a fully covariant way. We will explore the explicit form of $\hat R^{(0)}$ in the next section.
\section{Applications}
In this section, we explore three direct applications of the proposed formalism. First, we derive the finite non-relativistic limit of the two-derivative bosonic supergravity Lagrangian. We then discuss the convenience of this formulation for extracting covariant contributions at any desired order in the expansion in powers of $c$. As an illustrative example, we compute all finite contributions arising from the Metsaev--Tseytlin Lagrangian, which constitutes part of the four-derivative corrections to the bosonic Lagrangian in the NR limit%
\footnote{Additional contributions are expected to arise in order to cancel the higher-derivative divergences of the action.}. 
Finally, we discuss the possibility of employing more general non-metricities, focusing in particular on the case in which all such tensors vanish. This choice explicitly breaks the boost symmetry of the NR limit of bosonic supergravity.

\subsection{NR-Bosonic supergravity Lagrangian: metric-like (re-)formulation}

As we mentioned in the introduction, all the divergences of $L^{(2)}$ are controlled by the field $c_{\mu \nu}$, which in this pure-metric formulation 
requires the following constraint
\bea
\nabla_{[\mu|}{c_{\nu |\rho]}} \tau^{\nu} = \nabla_{[\mu}{\tau_{\rho]}} \, .
\label{extra}
\eea
Let us explicitly review the cancellation. The divergent contributions obtained from the Ricci scalar are given by
\bea
\hat R^{(2)} = - \frac12 \partial_{\mu}{\tau_{\nu}} \partial_{\rho}{\tau_{\sigma}} h^{\mu \sigma} h^{\nu \rho} + \frac12 \partial_{\mu}{\tau_{\nu}} \partial_{\rho}{\tau_{\sigma}} h^{\mu \rho} h^{\nu \sigma} \, ,
\eea
and the contributions from the divergent part of the $- \frac{1}{12}\hat H^2$ term is given by
\bea
- \frac12 \nabla_{\mu}{c_{\nu \rho}} \nabla_{\sigma}{c_{\gamma \epsilon}} \tau^{\nu} \tau^{\gamma} c c h^{\mu \sigma} h^{\rho \epsilon} + \frac12 \nabla_{\mu}{c_{\nu \rho}} \nabla_{\sigma}{c_{\gamma \epsilon}} \tau^{\nu} \tau^{\gamma} c c h^{\mu \epsilon} h^{\rho \sigma} \, .
\eea
By recalling that $\partial_{[\mu} \tau_{\nu]} = \nabla_{[\mu} \tau_{\nu]}$ and using (\ref{extra}) we obtain the cancellation. On the other hand, the finite Lagrangian contains contributions coming from the Ricci scalar, the dilaton term, and the $\hat H$-terms, and using the results of the previous section it can be written in the following way,
\bea
e^{-1} e^{2 \varphi} L^{(0)}|_{NR} & = & R_{\mu \nu}(\tau,h) h^{\mu \nu} +  h^{\mu \sigma} \tau^{\alpha \beta} \nabla_{\alpha} \tau_{[\mu \sigma}
(\nabla_{\nu]} h_{\epsilon \beta} - \frac12 \nabla_{\beta} h_{\nu] \epsilon}) h^{\epsilon \nu} 
\nn \\ && 
+  \tau^{\mu \sigma} h^{\alpha \beta} \nabla_{\alpha} h_{[\mu \sigma}  
(\nabla_{\epsilon} \tau_{\nu] \beta} - \frac12 \nabla_{\beta} \tau_{\nu] \epsilon}) h^{\epsilon \nu} 
 + 4 \nabla_{[\beta} \big[ h^{\beta \alpha}(  \nabla_{\mu} \tau_{\nu] \alpha} - \frac12 \nabla_{\alpha} \tau_{\nu] \mu} \big] \tau^{\mu \nu} \nn \\ && + 4 h^{\mu \nu} \nabla_{\mu} \varphi \nabla_{\nu} \varphi  - \frac{1}{12} h_{\mu \nu \rho} h_{\sigma \gamma \epsilon} h^{\mu \sigma} h^{\nu \gamma} h^{\rho \epsilon} - \frac12 h_{\mu \nu \rho} C_{\sigma \gamma \epsilon} \tau^{\mu \sigma} h^{\nu \gamma} h^{\rho \epsilon} \nn \\ && - \frac14 C_{\mu \nu \rho} C_{\sigma \gamma \epsilon} \tau^{\mu \sigma} \tau^{\nu \gamma} h^{\rho \epsilon} \, ,
\eea
where
\bea
 h_{\mu \nu \rho} & = & 3 \partial_{[\mu} b_{\nu \rho]} \, , \quad C_{\mu \nu \rho} =  - 3  \nabla_{[\mu} c_{\nu \rho]} \, . \nn
\eea

Now we can see that the finite bosonic Lagrangian after taking the NR limit contains a $R_{\mu \nu} h^{\mu \nu}$ contribution, plus other terms related to the non-metricities. As we mentioned before, the construction is purely written in metric formalism and is equivalent to the torsionful-formalism of string Newton-Cartan in the sense that no geometric constraints are imposed at the Lagrangian level. The non-metricities are just a convenient way of rewriting diffeomorphism-covariant quantities. Moreover, one cannot consider them as true non-metricities, in the sense that the fields $\tau$ and $h$ are not true metrics in the Newton-Cartan geometry. It is worth mentioning that the compatibility explored in this paper matches at the level of the action. The compatibility at the equations of motion is not guaranteed, since one should find a metric constraint equivalent to the torsion constraint found in \cite{NSNS}.  

\subsection{Non-relativistic bosonic $\alpha'$-corrections}
The four-derivative corrections to the bosonic string supergravity were historically computed considering three- and four-point scattering amplitudes for the massless states \cite{GrossSloan}-\cite{CN} (see \cite{Electure} for a pedagogical review of this topic). The method is based on the study of the different types of string interactions through the S-matrix to construct an effective Lagrangian, originally computed by Metsaev and Tseytlin\cite{MetsaevTseytlin},
\bea
S_{MT} = \int d^{26}x \sqrt{-g} e^{-2\phi} (L^{(0)} + L_{\rm MT}) \, ,
\label{fullA}
\eea
where 
\bea
L_{\rm MT} & = & - \frac{\alpha'}{4} \Big[  \hat R_{\mu \nu \rho \sigma} \hat R^{\mu \nu \rho \sigma} - \frac12 \hat H^{\mu \nu \rho} \hat H_{\mu \sigma \lambda} \hat R_{\nu \rho}{}^{\sigma \lambda}  + \frac{1}{24} \hat H^4 
- \frac18 \hat H^2_{\mu \nu} \hat H^{2 \mu \nu} \Big] \, ,  \, 
\label{MT}
\eea
and
\bea
\label{CSdef}
\hat H^2_{\mu \nu} & = & \hat H_{\mu}{}^{\rho \sigma} \hat H_{\nu \rho \sigma} \, , \quad \hat H^2  = \hat H_{\mu \nu \rho} \hat H^{\mu \nu \rho} \, .
\eea
The formalism developed in this work allows one to easily inspect the NR contributions of the previous quantities without losing covariance.

\subsubsection{Divergences of the four-derivative bosonic action}
Recalling that $\alpha' \rightarrow \frac{\alpha'_{NR}}{c^2}$ \cite{Higher-Derivative3}-\cite{Higher-Derivative4}, the divergent contributions coming from 
the Metsaev and Tseytlin Lagrangian are given by
\bea
(\hat R_{\mu \nu \rho \sigma} \hat R^{\mu \nu \rho \sigma})^{(4)}  & = & \hat R^{(2)\mu}{}_{\nu \rho \sigma} \hat R^{(2)\gamma}{}_{\epsilon \delta \lambda}  h_{\mu \gamma} h^{\nu \epsilon} h^{\rho \delta} h^{\sigma \lambda} + \hat R^{(2)\mu}{}_{\nu \rho \sigma} \hat R^{(2)\gamma}{}_{\epsilon \delta \lambda} \tau_{\mu} \tau_{\gamma} \tau^{\sigma} \tau^{\lambda}  h^{\nu \epsilon} h^{\rho \delta} \nn \\ && + \hat R^{(2)\mu}{}_{\nu \rho \sigma} \hat R^{(2)\gamma}{}_{\epsilon \delta \lambda} \tau_{\mu} \tau_{\gamma} \tau^{\rho} \tau^{\delta}  h^{\nu \epsilon} h^{\sigma \lambda}  + \hat R^{(2)\mu}{}_{\nu \rho \sigma} \hat R^{(2)\gamma}{}_{\epsilon \delta \lambda} \tau_{\mu} \tau_{\gamma} \tau^{\nu} \tau^{\epsilon}  h^{\rho \delta} h^{\sigma \lambda} \nn \\ && + 2 \hat R^{(0)\mu}{}_{\nu \rho \sigma} \hat R^{(2)\gamma}{}_{\epsilon \delta \lambda} \tau_{\mu} \tau_{\gamma} h^{\nu \epsilon} h^{\rho \delta} h^{\sigma \lambda}  \nn \\
& = & \frac32 \partial_{\mu}{\tau_{\nu}} \partial_{\rho}{\tau_{\sigma}} \partial_{\gamma}{\tau_{\epsilon}} \partial_{\delta}{\tau_{\lambda}}  h^{\mu \rho} h^{\nu \sigma} h^{\gamma \delta} h^{\epsilon \lambda} - 3 \partial_{\mu}{\tau_{\nu}} \partial_{\rho}{\tau_{\sigma}} \partial_{\gamma}{\tau_{\epsilon}} \partial_{\delta}{\tau_{\lambda}}  h^{\mu \rho} h^{\nu \sigma} h^{\gamma \lambda} h^{\epsilon \delta} \nn \\ && - 4 \partial_{\mu}{\tau_{\nu}} \partial_{\rho}{\tau_{\sigma}} \partial_{\gamma}{\tau_{\epsilon}} \partial_{\delta}{\tau_{\lambda}}  h^{\mu \rho} h^{\nu \gamma} h^{\sigma \lambda} h^{\epsilon \delta} + 2 \partial_{\mu}{\tau_{\nu}} \partial_{\rho}{\tau_{\sigma}} \partial_{\gamma}{\tau_{\epsilon}} \partial_{\delta}{\tau_{\lambda}}  h^{\mu \rho} h^{\nu \gamma} h^{\sigma \delta} h^{\epsilon \lambda} \nn \\ && + \partial_{\mu}{\tau_{\nu}} \partial_{\rho}{\tau_{\sigma}} \partial_{\gamma}{\tau_{\epsilon}} \partial_{\delta}{\tau_{\lambda}}  h^{\mu \rho} h^{\nu \epsilon} h^{\sigma \lambda} h^{\gamma \delta} + \frac32 \partial_{\mu}{\tau_{\nu}} \partial_{\rho}{\tau_{\sigma}} \partial_{\gamma}{\tau_{\epsilon}} \partial_{\delta}{\tau_{\lambda}}  h^{\mu \sigma} h^{\nu \rho} h^{\gamma \lambda} h^{\epsilon \delta} \nn \\ && + \partial_{\mu}{\tau_{\nu}} \partial_{\rho}{\tau_{\sigma}} \partial_{\gamma}{\tau_{\epsilon}} \partial_{\delta}{\tau_{\lambda}}  h^{\mu \sigma} h^{\nu \gamma} h^{\rho \lambda} h^{\epsilon \delta} \, , \\ 
- \frac12 (\hat H^{\mu \nu \rho} \hat H_{\mu \sigma \lambda} \hat R_{\nu \rho}{}^{\sigma \lambda})^{(4)} & = & 0 \, ,
\\
\frac{1}{24} (\hat H^4)^{(4)} & = & \frac32 \partial_{\mu}{c_{\nu \rho}} \partial_{\sigma}{c_{\gamma \epsilon}} \partial_{\delta}{c_{\lambda \alpha}} \partial_{\beta}{c_{\xi \chi}} \tau^{\nu} \tau^{\gamma} \tau^{\lambda} \tau^{\xi}  h^{\mu \sigma} h^{\rho \epsilon} h^{\delta \beta} h^{\alpha \chi} \nn \\ && - 3 \partial_{\mu}{c_{\nu \rho}} \partial_{\sigma}{c_{\gamma \epsilon}} \partial_{\delta}{c_{\lambda \alpha}} \partial_{\beta}{c_{\xi \chi}} \tau^{\nu} \tau^{\gamma} \tau^{\lambda} \tau^{\xi}  h^{\mu \sigma} h^{\rho \epsilon} h^{\delta \chi} h^{\alpha \beta} \nn \\ && + \frac32 \partial_{\mu}{c_{\nu \rho}} \partial_{\sigma}{c_{\gamma \epsilon}} \partial_{\delta}{c_{\lambda \alpha}} \partial_{\beta}{c_{\xi \chi}} \tau^{\nu} \tau^{\gamma} \tau^{\lambda} \tau^{\xi}  h^{\mu \epsilon} h^{\rho \sigma} h^{\delta \chi} h^{\alpha \beta} \, , \\
- \frac18 (\hat H^2_{\mu \nu} \hat H^{2 \mu \nu})^{(4)} & = & - \partial_{\mu}{c_{\nu \rho}} \partial_{\sigma}{c_{\gamma \epsilon}} \partial_{\delta}{c_{\lambda \alpha}} \partial_{\beta}{c_{\xi \chi}} \tau^{\nu} \tau^{\gamma} \tau^{\lambda} \tau^{\xi}  h^{\mu \sigma} h^{\rho \alpha} h^{\epsilon \chi} h^{\delta \beta} \nn \\ && + 4 \partial_{\mu}{c_{\nu \rho}} \partial_{\sigma}{c_{\gamma \epsilon}} \partial_{\delta}{c_{\lambda \alpha}} \partial_{\beta}{c_{\xi \chi}} \tau^{\nu} \tau^{\gamma} \tau^{\lambda} \tau^{\xi}  h^{\mu \sigma} h^{\rho \delta} h^{\epsilon \chi} h^{\alpha \beta} \nn \\ && - 2 \partial_{\mu}{c_{\nu \rho}} \partial_{\sigma}{c_{\gamma \epsilon}} \partial_{\delta}{c_{\lambda \alpha}} \partial_{\beta}{c_{\xi \chi}} \tau^{\nu} \tau^{\gamma} \tau^{\lambda} \tau^{\xi}  h^{\mu \sigma} h^{\rho \delta} h^{\epsilon \beta} h^{\alpha \chi} \nn \\ && - \partial_{\mu}{c_{\nu \rho}} \partial_{\sigma}{c_{\gamma \epsilon}} \partial_{\delta}{c_{\lambda \alpha}} \partial_{\beta}{c_{\xi \chi}} \tau^{\nu} \tau^{\gamma} \tau^{\lambda} \tau^{\xi}  h^{\mu \epsilon} h^{\rho \delta} h^{\sigma \chi} h^{\alpha \beta} \nn \\ && - \frac12 \partial_{\mu}{c_{\nu \rho}} \partial_{\sigma}{c_{\gamma \epsilon}} \partial_{\delta}{c_{\lambda \alpha}} \partial_{\beta}{c_{\xi \chi}} \tau^{\nu} \tau^{\gamma} \tau^{\lambda} \tau^{\xi}  h^{\mu \sigma} h^{\rho \epsilon} h^{\delta \beta} h^{\alpha \chi} \nn \\ && + \partial_{\mu}{c_{\nu \rho}} \partial_{\sigma}{c_{\gamma \epsilon}} \partial_{\delta}{c_{\lambda \alpha}} \partial_{\beta}{c_{\xi \chi}} \tau^{\nu} \tau^{\gamma} \tau^{\lambda} \tau^{\xi}  h^{\mu \sigma} h^{\rho \epsilon} h^{\delta \chi} h^{\alpha \beta} \nn \\ && - \frac12 \partial_{\mu}{c_{\nu \rho}} \partial_{\sigma}{c_{\gamma \epsilon}} \partial_{\delta}{c_{\lambda \alpha}} \partial_{\beta}{c_{\xi \chi}} \tau^{\nu} \tau^{\gamma} \tau^{\lambda} \tau^{\xi}  h^{\mu \epsilon} h^{\rho \sigma} h^{\delta \chi} h^{\alpha \beta} \, .
\eea
After imposing the condition (\ref{extra}) the previous expressions simplify to finally obtain
\bea
L^{\rm div}|_{\rm MT} & = & - \frac{5}{8}c^{2}\alpha'_{NR} \Big[ \nabla_{\mu}{\tau_{\nu}} \nabla_{\rho}{\tau_{\sigma}} \nabla_{\gamma}{\tau_{\epsilon}} \nabla_{\delta}{\tau_{\lambda}}  h^{\mu \rho} h^{\nu \sigma} h^{\gamma \delta} h^{\epsilon \lambda} \nn \\ && - 2 \nabla_{\mu}{\tau_{\nu}} \nabla_{\rho}{\tau_{\sigma}} \nabla_{\gamma}{\tau_{\epsilon}} \nabla_{\delta}{\tau_{\lambda}}  h^{\mu \rho} h^{\nu \sigma} h^{\gamma \lambda} h^{\epsilon \delta} \nn \\ && + \nabla_{\mu}{\tau_{\nu}} \nabla_{\rho}{\tau_{\sigma}} \nabla_{\gamma}{\tau_{\epsilon}} \nabla_{\delta}{\tau_{\lambda}}  h^{\mu \sigma} h^{\nu \rho} h^{\gamma \lambda} h^{\epsilon \delta} \Big] \, .
\label{divergenceMT}
\eea
The previous terms survive, showing that on the Metsaev and Tseytlin action the limit $c\rightarrow\infty$ leads to divergences. However, it is expected that they could be absorbed in field redefinitions and work in progress is being made towards this proof \cite{LRZ}. Another possibility is to impose the constraint
\bea
\label{const}
h^{\mu \nu} \nabla_{\mu} \tau_{\rho} = 0 \, ,
\eea
which cancels the divergences.

\subsubsection{Finite contributions to the four-derivative action}
The finite contributions of the four-derivative action can also be easily computed with the formalism here developed. The finite Riemann-squared contributions are given by
\bea
&& 2 \hat R^{(4)\mu}{}_{\nu \rho \sigma} \hat R^{(-4)\gamma}{}_{\epsilon \delta \lambda} \tau_{\mu \gamma} h^{\nu \epsilon} h^{\rho \delta} h^{\sigma \lambda} + 2 \hat R^{(2)\mu}{}_{\nu \rho \sigma} \hat R^{(-2)\gamma}{}_{\epsilon \delta \lambda} \tau_{\mu \gamma} h^{\nu \epsilon} h^{\rho \delta} h^{\sigma \lambda} + 2 \hat R^{(4)\mu}{}_{\nu \rho \sigma} \hat R^{(-2)\gamma}{}_{\epsilon \delta \lambda} h_{\mu \gamma} h^{\nu \epsilon} h^{\rho \delta} h^{\sigma \lambda} \nn \\ && + 4 \hat R^{(4)\mu}{}_{\nu \rho \sigma} \hat R^{(-2)\gamma}{}_{\epsilon \delta \lambda} \tau_{\mu \gamma} \tau^{\rho \delta} h^{\nu \epsilon} h^{\sigma \lambda}  + 2 \hat R^{(4)\mu}{}_{\nu \rho \sigma} \hat R^{(-2)\gamma}{}_{\epsilon \delta \lambda} \tau_{\mu \gamma} \tau^{\nu \epsilon} h^{\rho \delta} h^{\sigma \lambda} + \hat R^{(0)\mu}{}_{\nu \rho \sigma} \hat R^{(0)\gamma}{}_{\epsilon \delta \lambda} \tau_{\mu \gamma} h^{\nu \epsilon} h^{\rho \delta} h^{\sigma \lambda} \nn \\ && + 2 \hat R^{(0)\mu}{}_{\nu \rho \sigma} \hat R^{(2)\gamma}{}_{\epsilon \delta \lambda} h_{\mu \gamma} h^{\nu \epsilon} h^{\rho \delta} h^{\sigma \lambda} + 4 \hat R^{(0)\mu}{}_{\nu \rho \sigma} \hat R^{(2)\gamma}{}_{\epsilon \delta \lambda} \tau_{\mu \gamma} \tau^{\rho \delta} h^{\nu \epsilon} h^{\sigma \lambda} + 2 \hat R^{(0)\mu}{}_{\nu \rho \sigma} \hat R^{(2)\gamma}{}_{\epsilon \delta \lambda} \tau_{\mu \gamma} \tau^{\nu \epsilon} h^{\rho \delta} h^{\sigma \lambda} \nn \\ && + 4 \hat R^{(0)\mu}{}_{\nu \rho \sigma} \hat R^{(4)\gamma}{}_{\epsilon \delta \lambda} \tau^{\rho \delta} h_{\mu \gamma} h^{\nu \epsilon} h^{\sigma \lambda} + 2 \hat R^{(0)\mu}{}_{\nu \rho \sigma} \hat R^{(4)\gamma}{}_{\epsilon \delta \lambda} \tau^{\nu \epsilon} h_{\mu \gamma} h^{\rho \delta} h^{\sigma \lambda} + 2 \hat R^{(0)\mu}{}_{\nu \rho \sigma} \hat R^{(4)\gamma}{}_{\epsilon \delta \lambda} \tau_{\mu \gamma} \tau^{\rho \delta} \tau^{\sigma \lambda} h^{\nu \epsilon} \nn \\ && + 4 \hat R^{(0)\mu}{}_{\nu \rho \sigma} \hat R^{(4)\gamma}{}_{\epsilon \delta \lambda} \tau_{\mu \gamma} \tau^{\nu \epsilon} \tau^{\rho \delta} h^{\sigma \lambda} + 2 \hat R^{(2)\mu}{}_{\nu \rho \sigma} \hat R^{(2)\gamma}{}_{\epsilon \delta \lambda} \tau^{\rho \delta} h_{\mu \gamma} h^{\nu \epsilon} h^{\sigma \lambda} + \hat R^{(2)\mu}{}_{\nu \rho \sigma} \hat R^{(2)\gamma}{}_{\epsilon \delta \lambda} \tau^{\nu \epsilon} h_{\mu \gamma} h^{\rho \delta} h^{\sigma \lambda} \nn \\ && + \hat R^{(2)\mu}{}_{\nu \rho \sigma} \hat R^{(2)\gamma}{}_{\epsilon \delta \lambda} \tau_{\mu \gamma} \tau^{\rho \delta} \tau^{\sigma \lambda} h^{\nu \epsilon} + 2 \hat R^{(2)\mu}{}_{\nu \rho \sigma} \hat R^{(2)\gamma}{}_{\epsilon \delta \lambda} \tau_{\mu \gamma} \tau^{\nu \epsilon} \tau^{\rho \delta} h^{\sigma \lambda} + 2 \hat R^{(2)\mu}{}_{\nu \rho \sigma} \hat R^{(4)\gamma}{}_{\epsilon \delta \lambda} \tau^{\rho \delta} \tau^{\sigma \lambda} h_{\mu \gamma} h^{\nu \epsilon} \nn \\ && + 4 \hat R^{(2)\mu}{}_{\nu \rho \sigma} \hat R^{(4)\gamma}{}_{\epsilon \delta \lambda} \tau^{\nu \epsilon} \tau^{\rho \delta} h_{\mu \gamma} h^{\sigma \lambda} + 2 \hat R^{(2)\mu}{}_{\nu \rho \sigma} \hat R^{(4)\gamma}{}_{\epsilon \delta \lambda} \tau_{\mu \gamma} \tau^{\nu \epsilon} \tau^{\rho \delta} \tau^{\sigma \lambda} + \hat R^{(4)\mu}{}_{\nu \rho \sigma} \hat R^{(4)\gamma}{}_{\epsilon \delta \lambda} \tau^{\nu \epsilon} \tau^{\rho \delta} \tau^{\sigma \lambda} h_{\mu \gamma} \, .
\eea
Due to the length of the expressions, we provide the remaining H-dependent terms in the Appendix. 

The framework constructed in this manuscript now allows one to explore field redefinition for the different Newton-Cartan fields. For example, since $h^{\mu \nu}$ is boost invariant one can propose
\bea
\tilde h^{\mu \nu} = h^{\mu \nu} + a c^4 h^{\mu \rho} R^{(4)}_{\rho \sigma} h^{\sigma \nu} \, ,
\label{redefinitionhup}
\eea
which is a possible boost-invariant field redefinition for the $h^{\mu \nu}$ field since $\delta_{\lambda} R^{(4)}_{\mu \nu} = 0$.
\end{widetext}
The idea of the previous field redefinition is to induce divergences through the two-derivative Lagrangian by fixing the arbitrary coefficients $a$, which might cancel the divergences of the Metsaev and Tseytlin Lagrangian. Similar field redefinitions can be done for the rest of the fields, i.e., $h_{\mu \nu}$,$\tau_{\mu \nu}$, $\tau^{\mu \nu}$, $b_{\mu \nu}$ and $\phi$, consistently with the NR symmetries (one might expect deformations in the leading symmetries, coming from non-covariant field redefinitions). While the Metsaev and Tseytlin action can be written purely in terms of metric degrees of freedom, the boost invariance of the resulting action is not guaranteed due to the absence of the Stueckelberg field $m_{\mu}$ in the current formulation. Therefore, at four-derivative level one might expect new deformations to compensate the boost transformations \cite{NSNS}. 

A similar feature is expected to happen with the dilatation symmetry, since the missed Poisson equation is expected to have $\alpha'$-corrections. In this regard, we expects corrections to the dilation symmetry coming from field redefinitions.    

\subsection{General f(R,Q) non-relativistic theories}
While in this work we have shown that the non-metricities (\ref{Q1})-(\ref{Q4}) need to be fixed in a very particular way to provide the relativistic metric conditions (\ref{metriccond1})-(\ref{metriccond2}), one can construct more general NR theories with arbitrary non-metricities after taking the non-relativistic limit. One interesting example is to demand the compatibility condition on the Newton-Cartan fields, 
\bea
\nabla_{\mu} \tau_{\nu \rho} & = & 0 \, , \\
\nabla_{\mu} \tau^{\nu \rho} & = & 0  \, , \\
\nabla_{\mu} h_{\nu \rho} & = & 0 \, , \\
\nabla_{\mu} h^{\nu \rho} & = & 0\, , 
\eea
which automatically implies that both the relativistic Riemann and Ricci tensor are now finite,
\bea
\hat R^{\rho}{}_{\epsilon \mu \nu} & = & R^{\rho}{}_{\epsilon \mu \nu} \, , \quad
\hat R_{\epsilon \nu} = R_{\epsilon \nu} \, ,
\eea
and the more general (gravitational) Lagrangian in this setup is given by
\bea
L_{NR}|_{Q=0} = a_1 R_{\mu \nu} h^{\mu \nu} + a_2 R_{\mu \nu} \tau^{\mu \nu} \, ,
\eea
with $a_1$ and $a_2$ being two arbitrary coefficients. In this case, the $SO(8)$ and $SO(1,1)$ symmetry from the vielbein formalism can be preserved, but the boost symmetry is broken. Further theories with these properties can be explored by constructing generic $L(R,Q)$ Lagrangians in terms of $\tau$ and $h$, where a particular combination will be associated to the low energy limit of the bosonic supergravity. For example, it is well-known that the Lagrangian
\bea
L(\hat g, Q) & = & \sqrt{-g} \Big(c_1 Q_{\lambda} Q^{\lambda} + c_2 Q_{\lambda} P^{\lambda} + + c_3 P_{\lambda} P^{\lambda} \nn \\ && + c_4 Q^{\mu}{}_{\nu \lambda} Q_{\mu}{}^{\nu \lambda} + c_5 Q_{\nu}{}^{\mu}{}_{\lambda} Q_{\mu}{}^{\lambda \nu} \Big) \, ,
\eea
with $Q^{\mu}= Q^{\mu}{}_{\rho \sigma} \hat g^{\rho \sigma}$ and $P_{\nu}=Q^{\mu}{}_{\nu \mu}$ recovers the dynamics of GR when $c_1=-\frac14, c_2=\frac12$, $c_3=0$, $c_4=\frac14$, $c_5=-\frac12$, while more arbitrary combinations allow new diffeomorphism invariant Lagrangians. A similar setup is now possible to explore in the Newton-Cartan geometry. 

\section{Discussion}
The present construction provides a geometrical formulation of the non-relativistic limit of bosonic supergravity in terms of a Newton--Cartan geometry with non-metricities. Within this framework, an appropriate choice of connection allows a metric formulation, mimicking the construction of the relativistic cases. As a result, the full two-derivative bosonic supergravity in its NR limit can be written in a covariant form with respect to infinitesimal diffeomorphisms, thereby complementing other approaches based on subgroups of the Lorentz symmetry. Moreover, we find compatibility of the approaches at the level of the action, while the study of the equations of motion deserves further study to understand if it is necessary a new constraint, equivalent to the torsion constraint \cite{NSNS}.

A key advantage of the construction is that it allows for a straightforward decomposition of bosonic higher-derivative contributions, such as $\alpha'$-corrections. These corrections are naturally formulated in the metric formalism, where the Lorentz symmetry is manifest and no vielbein (or spin connection) is required. From a practical perspective, this provides a systematic method for analyzing higher-derivative terms by expressing relativistic curvature invariants directly in terms of non-relativistic geometric data. In particular, the use of fixed non-metricities enables a transparent identification of divergent and finite contributions, a task that is typically challenging.

Although we do not yet have full control over the divergences that appear at the four-derivative  (\ref{divergenceMT}), the formalism developed in this work already proves useful for isolating and organizing several finite contributions to the four-derivative bosonic Lagrangian in the NR limit. In particular, we have explicitly demonstrated how to compute the divergent and finite contributions arising from the Metsaev--Tseytlin Lagrangian. We expect that the covariant rewriting presented here will provide a valuable starting point for future investigations aimed at resolving these divergences in the bosonic $\alpha'$-corrections, as well as for extending the analysis to heterotic supergravity and more general non-relativistic gravity models.

\begin{widetext}
\acknowledgments
This work is supported by the SONATA BIS grant 2021/42/E/ST2/00304 from the National Science Centre (NCN), Poland. The author is very grateful to Jan Rosseel for discussions and clarifications on the article \cite{NSNS}.

\begin{appendix}
\section{Full finite Metsaev
and Tseytlin Lagrangian-H-dependent terms}
The terms only containing $\hat H$-contributions can also be easily written in covariant form. For example, the $\hat H^4$ term contributes with the following finite terms,
\bea
&& 4 \hat H^{(2)}_{\mu \nu \rho} h^{\mu \sigma} h^{\nu \gamma} h^{\rho \epsilon} h^{\delta \lambda} h^{\alpha \beta} h^{\chi\psi} h_{\sigma \gamma \epsilon} h_{\delta \alpha \chi} h_{\lambda \beta \psi} + 6 \hat H^{(2)}_{\mu \nu \rho} \hat H^{(2)}_{\sigma \gamma \epsilon} \tau^{\mu \sigma} h^{\nu \gamma} h^{\rho \epsilon} h^{\delta \lambda} h^{\alpha \beta} h^{\chi\psi} h_{\delta \alpha \chi} h_{\lambda \beta \psi} \nn \\ && + 6 \hat H^{(2)}_{\mu \nu \rho} \hat H^{(2)}_{\sigma \gamma \epsilon} \tau^{\delta \lambda} h^{\mu \sigma} h^{\nu \gamma} h^{\rho \epsilon} h^{\alpha \beta} h^{\chi\psi} h_{\delta \alpha \chi} h_{\lambda \beta \psi} + 24 \hat H^{(2)}_{\mu \nu \rho} \hat H^{(2)}_{\sigma \gamma \epsilon} \tau^{\mu \delta} h^{\nu \lambda} h^{\rho \alpha} h^{\sigma \beta} h^{\gamma \chi} h^{\epsilon \psi} h_{\delta \lambda \alpha} h_{\beta \chi\psi} \nn \\ && + 12 \hat H^{(2)}_{\mu \nu \rho} \hat H^{(2)}_{\sigma \gamma \epsilon} \hat H^{(2)}_{\delta \lambda \alpha} \tau^{\mu \sigma} \tau^{\nu \gamma} h^{\rho \epsilon} h^{\delta \beta} h^{\lambda \chi} h^{\alpha \psi} h_{\beta \chi\psi} + 36 \hat H^{(2)}_{\mu \nu \rho} \hat H^{(2)}_{\sigma \gamma \epsilon} \hat H^{(2)}_{\delta \lambda \alpha} \tau^{\mu \sigma} \tau^{\delta \beta} h^{\nu \gamma} h^{\rho \epsilon} h^{\lambda \chi} h^{\alpha \psi} h_{\beta \chi\psi} \nn \\ && + 12 \hat H^{(2)}_{\mu \nu \rho} \hat H^{(2)}_{\sigma \gamma \epsilon} \hat H^{(2)}_{\delta \lambda \alpha} \tau^{\mu \beta} \tau^{\nu \chi} h^{\rho \psi} h^{\sigma \delta} h^{\gamma \lambda} h^{\epsilon \alpha} h_{\beta \chi\psi} + 2 \hat H^{(2)}_{\mu \nu \rho} \hat H^{(2)}_{\sigma \gamma \epsilon} \hat H^{(2)}_{\delta \lambda \alpha} \hat H^{(2)}_{\beta \chi\psi} \tau^{\mu \sigma} \tau^{\nu \gamma} \tau^{\rho \epsilon} h^{\delta \beta} h^{\lambda \chi} h^{\alpha \psi} \nn \\ && + 18 \hat H^{(2)}_{\mu \nu \rho} \hat H^{(2)}_{\sigma \gamma \epsilon} \hat H^{(2)}_{\delta \lambda \alpha} \hat H^{(2)}_{\beta \chi\psi} \tau^{\mu \sigma} \tau^{\nu \gamma} \tau^{\delta \beta} h^{\rho \epsilon} h^{\lambda \chi} h^{\alpha \psi} \, ,
\eea
while $H^2_{\mu \nu} H^{2\mu \nu}$ contributes with,
\bea
&& 4 \hat H^{(2)}_{\mu \nu \rho} h^{\mu \sigma} h^{\nu \gamma} h^{\rho \epsilon} h^{\delta \lambda} h^{\alpha \beta} h^{\chi \psi} h_{\sigma \gamma \delta} h_{\epsilon \alpha \chi} h_{\lambda \beta \psi} + 4 \hat H^{(2)}_{\mu \nu \rho} \hat H^{(2)}_{\sigma \gamma \epsilon} \tau^{\mu \sigma} h^{\nu \gamma} h^{\rho \delta} h^{\epsilon \lambda} h^{\alpha \beta} h^{\chi \psi} h_{\delta \alpha \chi} h_{\lambda \beta \psi} \nn \\ && + 4 \hat H^{(2)}_{\mu \nu \rho} \hat H^{(2)}_{\sigma \gamma \epsilon} \tau^{\mu \delta} h^{\nu \sigma} h^{\rho \gamma} h^{\epsilon \lambda} h^{\alpha \beta} h^{\chi \psi} h_{\delta \alpha \chi} h_{\lambda \beta \psi} + 4 \hat H^{(2)}_{\mu \nu \rho} \hat H^{(2)}_{\sigma \gamma \epsilon} \tau^{\delta \lambda} h^{\mu \sigma} h^{\nu \gamma} h^{\rho \alpha} h^{\epsilon \beta} h^{\chi \psi} h_{\delta \alpha \chi} h_{\lambda \beta \psi} \nn \\ && - 8 \hat H^{(2)}_{\mu \nu \rho} \hat H^{(2)}_{\sigma \gamma \epsilon} \tau^{\mu \delta} h^{\nu \sigma} h^{\rho \lambda} h^{\gamma \alpha} h^{\epsilon \beta} h^{\chi \psi} h_{\delta \lambda \chi} h_{\alpha \beta \psi} + 2 \hat H^{(2)}_{\mu \nu \rho} \hat H^{(2)}_{\sigma \gamma \epsilon} \tau^{\mu \sigma} h^{\nu \delta} h^{\rho \lambda} h^{\gamma \alpha} h^{\epsilon \beta} h^{\chi \psi} h_{\delta \lambda \chi} h_{\alpha \beta \psi} \nn \\ && + 2 \hat H^{(2)}_{\mu \nu \rho} \hat H^{(2)}_{\sigma \gamma \epsilon} \tau^{\delta \lambda} h^{\mu \sigma} h^{\nu \alpha} h^{\rho \beta} h^{\gamma \chi} h^{\epsilon \psi} h_{\delta \alpha \beta} h_{\lambda \chi \psi} + 8 \hat H^{(2)}_{\mu \nu \rho} \hat H^{(2)}_{\sigma \gamma \epsilon} \tau^{\mu \delta} h^{\nu \lambda} h^{\rho \alpha} h^{\sigma \beta} h^{\gamma \chi} h^{\epsilon \psi} h_{\delta \lambda \beta} h_{\alpha \chi \psi} \nn \\ && + 4 \hat H^{(2)}_{\mu \nu \rho} \hat H^{(2)}_{\sigma \gamma \epsilon} \tau^{\mu \delta} h^{\nu \lambda} h^{\rho \alpha} h^{\sigma \beta} h^{\gamma \chi} h^{\epsilon \psi} h_{\delta \beta \chi} h_{\lambda \alpha \psi} + 4 \hat H^{(2)}_{\mu \nu \rho} \hat H^{(2)}_{\sigma \gamma \epsilon} \hat H^{(2)}_{\delta \lambda \alpha} \tau^{\mu \sigma} \tau^{\nu \gamma} h^{\rho \delta} h^{\epsilon \beta} h^{\lambda \chi} h^{\alpha \psi} h_{\beta \chi \psi} \nn \\ && - 8 \hat H^{(2)}_{\mu \nu \rho} \hat H^{(2)}_{\sigma \gamma \epsilon} \hat H^{(2)}_{\delta \lambda \alpha} \tau^{\mu \sigma} \tau^{\nu \delta} h^{\rho \gamma} h^{\epsilon \beta} h^{\lambda \chi} h^{\alpha \psi} h_{\beta \chi \psi} + 16 \hat H^{(2)}_{\mu \nu \rho} \hat H^{(2)}_{\sigma \gamma \epsilon} \hat H^{(2)}_{\delta \lambda \alpha} \tau^{\mu \sigma} \tau^{\delta \beta} h^{\nu \gamma} h^{\rho \lambda} h^{\epsilon \chi} h^{\alpha \psi} h_{\beta \chi \psi} \nn \\ && + 8 \hat H^{(2)}_{\mu \nu \rho} \hat H^{(2)}_{\sigma \gamma \epsilon} \hat H^{(2)}_{\delta \lambda \alpha} \tau^{\mu \sigma} \tau^{\nu \beta} h^{\rho \chi} h^{\gamma \delta} h^{\epsilon \lambda} h^{\alpha \psi} h_{\beta \chi \psi} + 4 \hat H^{(2)}_{\mu \nu \rho} \hat H^{(2)}_{\sigma \gamma \epsilon} \hat H^{(2)}_{\delta \lambda \alpha} \tau^{\mu \beta} \tau^{\nu \chi} h^{\rho \sigma} h^{\gamma \delta} h^{\epsilon \lambda} h^{\alpha \psi} h_{\beta \chi \psi} \nn \\ && - 8 \hat H^{(2)}_{\mu \nu \rho} \hat H^{(2)}_{\sigma \gamma \epsilon} \hat H^{(2)}_{\delta \lambda \alpha} \tau^{\mu \sigma} \tau^{\nu \beta} h^{\rho \gamma} h^{\epsilon \delta} h^{\lambda \chi} h^{\alpha \psi} h_{\beta \chi \psi} + 4 \hat H^{(2)}_{\mu \nu \rho} \hat H^{(2)}_{\sigma \gamma \epsilon} \hat H^{(2)}_{\delta \lambda \alpha} \tau^{\mu \sigma} \tau^{\delta \beta} h^{\nu \lambda} h^{\rho \alpha} h^{\gamma \chi} h^{\epsilon \psi} h_{\beta \chi \psi} \nn \\ && - 8 \hat H^{(2)}_{\mu \nu \rho} \hat H^{(2)}_{\sigma \gamma \epsilon} \hat H^{(2)}_{\delta \lambda \alpha} \tau^{\mu \beta} \tau^{\sigma \chi} h^{\nu \delta} h^{\rho \lambda} h^{\gamma \alpha} h^{\epsilon \psi} h_{\beta \chi \psi} + 4 \hat H^{(2)}_{\mu \nu \rho} \hat H^{(2)}_{\sigma \gamma \epsilon} \hat H^{(2)}_{\delta \lambda \alpha} \hat H^{(2)}_{\beta \chi \psi} \tau^{\mu \sigma} \tau^{\nu \gamma} \tau^{\rho \delta} h^{\epsilon \beta} h^{\lambda \chi} h^{\alpha \psi} \nn \\ && + 4 \hat H^{(2)}_{\mu \nu \rho} \hat H^{(2)}_{\sigma \gamma \epsilon} \hat H^{(2)}_{\delta \lambda \alpha} \hat H^{(2)}_{\beta \chi \psi} \tau^{\mu \sigma} \tau^{\nu \gamma} \tau^{\delta \beta} h^{\rho \lambda} h^{\epsilon \chi} h^{\alpha \psi} + 8 \hat H^{(2)}_{\mu \nu \rho} \hat H^{(2)}_{\sigma \gamma \epsilon} \hat H^{(2)}_{\delta \lambda \alpha} \hat H^{(2)}_{\beta \chi \psi} \tau^{\mu \sigma} \tau^{\nu \delta} \tau^{\gamma \beta} h^{\rho \lambda} h^{\epsilon \chi} h^{\alpha \psi} \nn \\ && + 4 \hat H^{(2)}_{\mu \nu \rho} \hat H^{(2)}_{\sigma \gamma \epsilon} \hat H^{(2)}_{\delta \lambda \alpha} \hat H^{(2)}_{\beta \chi \psi} \tau^{\mu \sigma} \tau^{\nu \delta} \tau^{\gamma \beta} h^{\rho \epsilon} h^{\lambda \chi} h^{\alpha \psi} \, .
\eea
For $\hat H^{\mu \nu \rho} \hat H_{\mu \sigma \lambda} \hat R_{\nu \rho}{}^{\sigma \lambda}$ the finite contributions are given by
\bea
&& \hat R^{(2)\mu}{}_{\nu \rho \sigma} h^{\nu \gamma} h^{\rho \epsilon} h^{\sigma \delta} h^{\lambda \alpha} h_{\mu \gamma \lambda} h_{\epsilon \delta \alpha} - \hat R^{(4)\mu}{}_{\nu \rho \sigma} \tau^{\gamma \epsilon} h^{\nu \delta} h^{\rho \lambda} h^{\sigma \alpha} h_{\mu \gamma \delta} h_{\epsilon \lambda \alpha} + 2 \hat R^{(4)\mu}{}_{\nu \rho \sigma} \tau^{\rho \gamma} h^{\nu \epsilon} h^{\sigma \delta} h^{\lambda \alpha} h_{\mu \epsilon \lambda} h_{\gamma \delta \alpha} \nn \\ && + \hat R^{(4)\mu}{}_{\nu \rho \sigma} \tau^{\nu \gamma} h^{\rho \epsilon} h^{\sigma \delta} h^{\lambda \alpha} h_{\mu \gamma \lambda} h_{\epsilon \delta \alpha} - \hat H^{(2)}_{\mu \nu \rho} \hat R^{(0)\sigma}{}_{\gamma \epsilon \delta} h^{\mu \epsilon} h^{\nu \delta} h^{\rho \lambda} h^{\gamma \alpha} h_{\sigma \lambda \alpha} - \hat H^{(2)}_{\mu \nu \rho} \hat R^{(2)\sigma}{}_{\gamma \epsilon \delta} \tau^{\mu \lambda} h^{\nu \epsilon} h^{\rho \delta} h^{\gamma \alpha} h_{\sigma \lambda \alpha} \nn \\ && - 2 \hat H^{(2)}_{\mu \nu \rho} \hat R^{(2)\sigma}{}_{\gamma \epsilon \delta} \tau^{\mu \epsilon} h^{\nu \delta} h^{\rho \lambda} h^{\gamma \alpha} h_{\sigma \lambda \alpha} + \hat H^{(2)}_{\mu \nu \rho} \hat R^{(2)\sigma}{}_{\gamma \epsilon \delta} \tau^{\gamma \lambda} h^{\mu \epsilon} h^{\nu \delta} h^{\rho \alpha} h_{\sigma \lambda \alpha} + 2 \hat H^{(2)}_{\mu \nu \rho} \hat R^{(4)\sigma}{}_{\gamma \epsilon \delta} \tau^{\mu \epsilon} \tau^{\nu \lambda} h^{\rho \delta} h^{\gamma \alpha} h_{\sigma \lambda \alpha} \nn \\ && - \hat H^{(2)}_{\mu \nu \rho} \hat R^{(4)\sigma}{}_{\gamma \epsilon \delta} \tau^{\mu \epsilon} \tau^{\nu \delta} h^{\rho \lambda} h^{\gamma \alpha} h_{\sigma \lambda \alpha} - \hat H^{(2)}_{\mu \nu \rho} \hat R^{(4)\sigma}{}_{\gamma \epsilon \delta} \tau^{\mu \lambda} \tau^{\gamma \alpha} h^{\nu \epsilon} h^{\rho \delta} h_{\sigma \lambda \alpha} + 2 \hat H^{(2)}_{\mu \nu \rho} \hat R^{(4)\sigma}{}_{\gamma \epsilon \delta} \tau^{\mu \epsilon} \tau^{\gamma \lambda} h^{\nu \delta} h^{\rho \alpha} h_{\sigma \lambda \alpha} \nn \\ && + \hat H^{(2)}_{\mu \nu \rho} \hat R^{(0)\mu}{}_{\sigma \gamma \epsilon} h^{\nu \sigma} h^{\rho \delta} h^{\gamma \lambda} h^{\epsilon \alpha} h_{\delta \lambda \alpha} - \hat H^{(2)}_{\mu \nu \rho} \hat R^{(2)\mu}{}_{\sigma \gamma \epsilon} \tau^{\nu \delta} h^{\rho \sigma} h^{\gamma \lambda} h^{\epsilon \alpha} h_{\delta \lambda \alpha} - 2 \hat H^{(2)}_{\mu \nu \rho} \hat R^{(2)\mu}{}_{\sigma \gamma \epsilon} \tau^{\gamma \delta} h^{\nu \sigma} h^{\rho \lambda} h^{\epsilon \alpha} h_{\delta \lambda \alpha} \nn \\ && + \hat H^{(2)}_{\mu \nu \rho} \hat R^{(2)\mu}{}_{\sigma \gamma \epsilon} \tau^{\nu \sigma} h^{\rho \delta} h^{\gamma \lambda} h^{\epsilon \alpha} h_{\delta \lambda \alpha} - 2 \hat H^{(2)}_{\mu \nu \rho} \hat R^{(4)\mu}{}_{\sigma \gamma \epsilon} \tau^{\nu \delta} \tau^{\gamma \lambda} h^{\rho \sigma} h^{\epsilon \alpha} h_{\delta \lambda \alpha} + \hat H^{(2)}_{\mu \nu \rho} \hat R^{(4)\mu}{}_{\sigma \gamma \epsilon} \tau^{\gamma \delta} \tau^{\epsilon \lambda} h^{\nu \sigma} h^{\rho \alpha} h_{\delta \lambda \alpha} \nn \\ && + \hat H^{(2)}_{\mu \nu \rho} \hat R^{(4)\mu}{}_{\sigma \gamma \epsilon} \tau^{\nu \sigma} \tau^{\rho \delta} h^{\gamma \lambda} h^{\epsilon \alpha} h_{\delta \lambda \alpha} - 2 \hat H^{(2)}_{\mu \nu \rho} \hat R^{(4)\mu}{}_{\sigma \gamma \epsilon} \tau^{\nu \sigma} \tau^{\gamma \delta} h^{\rho \lambda} h^{\epsilon \alpha} h_{\delta \lambda \alpha} - \hat H^{(2)}_{\mu \nu \rho} \hat H^{(2)}_{\sigma \gamma \epsilon} \hat R^{(-2)\mu}{}_{\delta \lambda \alpha} h^{\nu \sigma} h^{\rho \delta} h^{\gamma \lambda} h^{\epsilon \alpha} \nn \\ && - \hat H^{(2)}_{\mu \nu \rho} \hat H^{(2)}_{\sigma \gamma \epsilon} \hat R^{(0)\mu}{}_{\delta \lambda \alpha} \tau^{\nu \sigma} h^{\rho \delta} h^{\gamma \lambda} h^{\epsilon \alpha} + 2 \hat H^{(2)}_{\mu \nu \rho} \hat H^{(2)}_{\sigma \gamma \epsilon} \hat R^{(0)\mu}{}_{\delta \lambda \alpha} \tau^{\sigma \lambda} h^{\nu \gamma} h^{\rho \delta} h^{\epsilon \alpha} + \hat H^{(2)}_{\mu \nu \rho} \hat H^{(2)}_{\sigma \gamma \epsilon} \hat R^{(0)\mu}{}_{\delta \lambda \alpha} \tau^{\nu \delta} h^{\rho \sigma} h^{\gamma \lambda} h^{\epsilon \alpha} \nn \\ && - 2 \hat H^{(2)}_{\mu \nu \rho} \hat H^{(2)}_{\sigma \gamma \epsilon} \hat R^{(2)\mu}{}_{\delta \lambda \alpha} \tau^{\nu \sigma} \tau^{\gamma \lambda} h^{\rho \delta} h^{\epsilon \alpha} - \hat H^{(2)}_{\mu \nu \rho} \hat H^{(2)}_{\sigma \gamma \epsilon} \hat R^{(2)\mu}{}_{\delta \lambda \alpha} \tau^{\sigma \lambda} \tau^{\gamma \alpha} h^{\nu \epsilon} h^{\rho \delta} - \hat H^{(2)}_{\mu \nu \rho} \hat H^{(2)}_{\sigma \gamma \epsilon} \hat R^{(2)\mu}{}_{\delta \lambda \alpha} \tau^{\nu \sigma} \tau^{\rho \delta} h^{\gamma \lambda} h^{\epsilon \alpha} \nn \\ && - 2 \hat H^{(2)}_{\mu \nu \rho} \hat H^{(2)}_{\sigma \gamma \epsilon} \hat R^{(2)\mu}{}_{\delta \lambda \alpha} \tau^{\nu \delta} \tau^{\sigma \lambda} h^{\rho \gamma} h^{\epsilon \alpha} - \hat H^{(2)}_{\mu \nu \rho} \hat H^{(2)}_{\sigma \gamma \epsilon} \hat R^{(4)\mu}{}_{\delta \lambda \alpha} \tau^{\nu \sigma} \tau^{\gamma \lambda} \tau^{\epsilon \alpha} h^{\rho \delta} - 2 \hat H^{(2)}_{\mu \nu \rho} \hat H^{(2)}_{\sigma \gamma \epsilon} \hat R^{(4)\mu}{}_{\delta \lambda \alpha} \tau^{\nu \sigma} \tau^{\rho \delta} \tau^{\gamma \lambda} h^{\epsilon \alpha} \nn \\ && + \hat H^{(2)}_{\mu \nu \rho} \hat H^{(2)}_{\sigma \gamma \epsilon} \hat R^{(4)\mu}{}_{\delta \lambda \alpha} \tau^{\nu \delta} \tau^{\sigma \lambda} \tau^{\gamma \alpha} h^{\rho \epsilon} \, .
\eea

At present, we cannot control all the divergences of the full Metsaev-Tseytlin Lagrangian (terms given in \ref{divergenceMT}) without imposing further constraints (such as \ref{const}). For that reason we are not giving an exhaustive analysis of the four-derivative contributions of this Lagrangian in its NR limit. However, the covariant framework developed in this work could be very useful for attacking this open problem and part of the finite action is given by the previous contributions.

\end{appendix}

\end{widetext}
\end{document}